\documentclass[11pt]{article}

\usepackage{graphicx,amsmath,amsfonts,amssymb,dcolumn,amsthm,slashed,bbm,xspace,pgffor,tikz,mathbbol,latexsym,enumerate,amsmath}
\usepackage{soul}
\usepackage[colorlinks,citecolor=red,urlcolor=cyan,linkcolor=blue]{hyperref}
\usepackage[utf8]{inputenc}
\usepackage[english]{babel}
\usepackage{lmodern}
\usepackage{microtype}
\usepackage{cite}

\numberwithin{equation}{section}

\begin{document}

\title{\textbf{Geometrodynamical description of two-dimensional electrodynamics}}

\author{\textbf{Rodrigo F.~Sobreiro}\thanks{rodrigo\_sobreiro@id.uff.br}\\\\
\textit{{\small UFF - Universidade Federal Fluminense, Instituto de F\'isica,}}\\
\textit{{\small Av. Litorânea, s/n, 24210-346, Niter\'oi, RJ, Brasil.}}}

\date{}
\maketitle

\begin{abstract}
Two-dimensional pure electrodynamics is mapped into two-dimensional gravity in the first order formalism at classical and quantum levels. Due to the fact that the degrees of freedom of these two theories do not match, we are enforced to introduce extra fields from the beginning. These fields are introduced through a BRST exact boundary term, so they are harmless to the physical content of the theory. The map between electromagnetism and gravity fields generate a non-trivial Jacobian, which brings extra features (but also harmless to the physical content of the gravity theory) at quantum level. 
\end{abstract}

\maketitle

\section{Introduction}\label{intro}

Two-dimensional electrodynamics is of great interest in Physics, mainly because of its usage in two-dimensional condensed matter systems, see \cite{CastroNeto:2007fxn,Farajollahpour:2019kwj,Cahangirov:2009kwj} and references therein - It is worth mentioning that two-dimensional classical electrodynamics is an exact solvable system \cite{Staruszkewicz:1967xxx,Bialynicki-Birula:1971akx,Kosyakov:1999np,Kosyakov:2007qc,Kosyakov:2007np}. But also as toy models used to probe more complicated systems. The latter statement is also valid for gravity, specially in the pursuing of a consistent quantization of gravity \cite{Polyakov:1987zb,Gross:1989vs,Witten:1989ig,Chamseddine:1989dm,Fukuma:1990jw,Kazama:1992ex}. In the present work we show how we can connect both theories.

Geometrodynamical descriptions of gauge systems appear in the literature quite often, see for instance \cite{Achucarro:1986uwr,Witten:1988hc,Obukhov:1998gx,Sobreiro:2007pn,Sobreiro:2011hb,Assimos:2013eua,Assimos:2019yln,Assimos:2021eok} and references therein. Our aim in the present work is to establish a consistent map relating two-dimensional electrodynamics and two-dimensional gravity in the first order formalism (described by the zweibein and the Lorentz connection as fundamental fields). The starting electromagnetic action we work is composed by the Maxwell term and the Chern-Pontryagin density. The main problem we have to face is that two-dimensional electromagnetism is a gauge theory with two degrees of freedom (physical and non-physical) while two-dimensional gravity in the first order formalism carries six degrees of freedom (physical and non-physical). Thence, a naive map between both theories, although possible, is ill defined. To account for the extra degrees of freedom at the gravity side, we introduce, at the electromagnetism side, a set of extra fields via an extra term in the action. The auxiliary term is constructed as a Becchi-Rouet-Stora-Tyutin (BRST) \cite{Becchi:1975nq,Tyutin:1975qk,Piguet:1995er} exact quantity in order to not mess with the physical degrees of freedom. Moreover, the auxiliary action is a boundary term. So we can safely say that this term should be ``twice harmless''. 

To actually perform the map we choose a non-trivial way to relate the fields in both theories. Although non-trivial, the map is well defined. At classical level, the gravity theory obtained is composed by the Einstein-Hilbert term which, in two-dimensions, coincides with the Gauss-Bonnet term, a cosmological constant term and another BRST exact boundary term with matter fields. At quantum level, because the map is nontrivial, a non-trivial Jacobian appears in the generating functional. Nevertheless, the Jacobian can be localized with the help of extra fields. The extra fields can also be collected in a BRST exact term, ensuring that the physical sector of the model is untouched.

This work is organized as follows. In Section ``Electrodynamics'' we construct the two-dimensional electromagnetic theory. In Section ``Lovelock-Cartan gravity'' we construct the two-dimensional gravity theory. Section ``Electrodynamics-Gravity equivalence'' is devoted to the map between both theories at classical level. The equivalence is then discussed at quantum level in Section ``Electrodynamics-gravity equivalence at quantum level''. Finally, our conclusions are displayed in Section ``Conclusions''.

\section{Electrodynamics}\label{ed}

The fundamental ingredient in electrodynamics is the gauge field 1-form $A=A_idx^i$ where $i,j,k,\ldots\in\{0,1\}$ and $dx^i$ refer to a two-dimensional manifold which we take, for simplicity, as the Euclidean spacetime $\mathbb{R}^2$. The field $A$ is a connection with respect to the gauge group $U(1)$. The gauge transformation of $A$ is
\begin{equation}
\delta A=d\alpha\;,\label{gt1}
\end{equation}
where $u(x)=\exp{[i\alpha(x)]}\in U(1)$. The gauge invariant field strength 2-form is given by
\begin{equation}
F=dA=\frac{1}{2}F_{ij}dx^idx^j\;.\label{fs1}
\end{equation}
The dual field strength $\ast F$ is obtained from the action of the Hodge dual $\ast$ on the field strength by means of
\begin{equation}
\ast F=\frac{1}{2}\epsilon^{ij}F_{ij}\;,\label{fs2}
\end{equation}
where $\epsilon^{ij}$ is the Levi-Civita tensor.

At this point, the action we consider is the most general second order action with the properties of being gauge invariant, local and polynomial in the gauge field and its derivatives, and power counting renormalizable, namely
\begin{equation}
S_{ed}=\int\left(\kappa F+\frac{1}{2}F\ast F\right)\;,\label{ed1}
\end{equation}
where the parameter $\kappa$ is a mass parameter. Moreover, although we are not considering a specific type of matter interaction, the fundamental electric charge $e$ is at our disposal. The mass dimensions of all (tensorial component) ingredients are given by $[A]=0$ and $[d]=[e]=[\kappa]=1$. The second term in action \eqref{ed1} is the usual Maxwell term while the first term is known as the topological Chern-Pontryagin density.

Although we are dealing with electrodynamics at classical level, BRST technology \cite{Piguet:1995er} will be evoked along the entire paper. Therefore, defining the Faddeev-Popov ghost field $c$, and $s$ as the BRST nilpotent operator, the BRST transformations read
\begin{eqnarray}
sA&=&dc\;\nonumber\\
sc&=&0\;.\label{brs1a}
\end{eqnarray}

The electromagnetic action \eqref{ed1} is clearly a BRST closed, but not exact, quantity. Thence, it belongs to the nontrivial sector of the BRST cohomology. Consequently, $S_{ed}$ carries physical information. The BRST symmetry is typically used to introduce a gauge fixing in the elimination of spurious degrees of freedom of $S_{ed}$. Nevertheless, we will not deal with gauge fixing issues in the present work and fix our attention exclusively to the action \eqref{ed1}.

\section{Lovelock-Cartan gravity}\label{2G}

The second theory we deal with is two-dimensional gravity in the first order formalism \cite{Utiyama:1956sy,Kibble:1961ba,Sciama:1964wt,Lovelock:1971yv,Mardones:1990qc,Zanelli:2005sa,Zanelli:2012zz}. It is constructed over a generic two-dimensional differential manifold $\mathrm{M}$ which is assumed to be Hausdorff, paracompact and orientable \cite{Nakahara:2003nw}. Over $\mathrm{M}$ we define two fundamental fields associated with the geometric properties of $\mathrm{M}$ itself, the zweibein 1-form $e^a=e^a_\mu dX^\mu$ and the connection 1-form $\omega^{ab}=\omega^{ab}_\mu dX^\mu$, with $X$ being a generic point in $\mathrm{M}$. The frame indexes $a,b,c,\ldots,h\in\{0,1\}$ are associated with coordinates at the tangent space $T_X(\mathrm{M})\sim T_X$ while world indexes $\mu,\nu,\alpha,\ldots\in\{0,1\}$ are associated with local space-time coordinates. The coordinated point $X\in\mathrm{M}$ is introduced in different notation of $x\in\mathbb{R}^2$ in order to avoid confusion with the formulation of electrodynamics in the previous section. The local isometries in tangent space are characterized by the Abelian gauge group $SO(2)$. The $SO(2)$ gauge transformations are typically characterized by
\begin{eqnarray}
\delta\omega^a_{\phantom{a}b}&=&d\alpha^a_{\phantom{a}b}\;,\nonumber\\
\delta e^a&=&\alpha^a_{\phantom{a}b}e^b\;,\label{gt2}
\end{eqnarray}
where $U(X)=\exp(i\alpha^{ab}L_{ab})\in SO(2)$. The generator of the two-dimensional orthogonal group is denoted by $L_{ab}$ which is anti-symmetric in $a$ and $b$.

The zweibein is known to have two essential properties. The first one is that it carries the metric properties of spacetime since one can derive the metric tensor from it by means of $g_{\mu\nu}=\eta_{ab}e_\mu^ae_\nu^b$. Second, that it provides the existence of local inertial frames at each point $X$ by means of $dX^a=e^a_\mu dX^\mu$, ensuring the validity of the equivalence principle \cite{Wald:1984rg,DeSabbata:1986sv,Misner:1973prb}. The coordinate system $X^a$ is identified with the tangent space and all possible local inertial frames are connected by the orthogonal gauge group. Thence, gravity is just a gauge theory with the extra principle that the gauge group is identified with the local isometry group.

Torsion and curvature 2-forms are, respectively, given by
\begin{eqnarray}
T^a&=&De^a\;\;=\;\;de^a+\omega^a_{\phantom{a}b}e^b\;,\nonumber\\
R^a_{\phantom{a}b}&=&d\omega^a_{\phantom{a}b}\;.\label{2forms1}
\end{eqnarray}
In addition, we have at our disposal two invariant objects: the Levi-Civita tensor $\epsilon_{ab}$ and the tangent space metric $\eta_{ab}$.

With the above ingredients at hand, we are able to construct the most general local and polynomial action, power-counting renormalizable, depending only on the fields and their first order derivatives and not explicitly depending on the metric. The result is the well known Mardones-Zanelli action \cite{Mardones:1990qc,Moritsch:1993eg,Zanelli:2005sa,Zanelli:2012zz} describing Lovelock-Cartan (LC) two-dimensional gravity,
\begin{equation}
S_{LC}=\frac{1}{8\pi G}\int\epsilon_{ab}\left(R^{ab}+\frac{\Lambda^2}{2}e^ae^b\right)\;,\label{grav1}
\end{equation}
where $G$ and $\Lambda$ are Newton and cosmological constants, respectively. The mass dimension of the relevant objects in the action \eqref{grav1} are $[\mathrm{d}]=[\omega^a_{\phantom{a}b}]=[\Lambda]=1$ and $[e^a]=[G]=0$. The last term at the action \eqref{grav1} is recognized as the cosmological constant term while the first one is the Einstein-Hilbert term. The Einstein-Hilbert term is topological in two dimensions because it coincides with the Gauss-Bonnet topological invariant.

Again, it is convenient at this point to establish the BRST symmetry of \eqref{grav1}. Defining the gravitational Faddeev-Popov ghost \cite{Baulieu:1984iw,Baulieu:1984pf,Moritsch:1993eg} by $c^{ab}$, the action \eqref{grav1} is invariant under
\begin{eqnarray}
s\omega^{ab}&=&dc^{ab}\;,\nonumber\\
se^a&=&c^a_{\phantom{a}b}e^b\;,\nonumber\\
sc^{ab}&=&0\;.\label{brst2a}
\end{eqnarray}
The action $S_{LC}$ carries physical information since it is a BRST closed but not an exact quantity. 

It is worth mentioning that, in the first order formalism, the vielbein and the spin connection are taken as independent fields. Any possible relation between them are derived from the (classical or quantum) field equations and symmetries of the theory in question. In the specific case of $S_{grav}$, the absence of torsion explicitly in the action may induce the wrong conclusion that torsion vanishes. In fact, in the action $S_{grav}$, only one term contributes to the field equations, the volume term. Thus, under minimization, $S_{grav}$ defines a Plateau problem while $\omega^{ab}$ remains undetermined (there is no field equation for the spin connection). This ambiguity leaves a freedom to fix the spin connection at will. Therefore, there is no obvious way to state that torsion vanishes.

\section{Electrodynamics-Gravity equivalence}\label{EQUIV}

In this section we discuss, at classical level, how we can connect the electromagnetic action \eqref{ed1} to the LC action \eqref{grav1}.

\subsection{Naive map}

Following \cite{Obukhov:1998gx}, we can introduce a 2-form auxiliary field $\theta$, with $[\theta]=1$, and rewrite action \eqref{ed1} as
\begin{equation}
   S_e=\int\;\left(\kappa F+\theta*F-\frac{1}{2}\theta\ast\theta\right)\;.\label{ed2}
\end{equation}
The auxiliary field is obviously gauge (and BRST) invariant. It is quite intuitive that this action can be mapped into \eqref{grav1} due to the homomorphism $U(1)\longmapsto SO(2)$. In fact, the following map between fields\footnote{The factors depending on $\mu$ are required in order to adjust the mass dimensions of the fields. The form of the mapping \eqref{map1} is not unique. In fact, the factor $\mu$ is a generic factor with mass dimension 1 constructed from $\kappa$ and $e$ (\emph{e.g.}, $\mu=e^q\kappa^mf{(e^p\kappa^{-p})}$ with $q+m=1$, $p\in\mathbb{R}$ and $f$ any analytical function). The target action \eqref{grav2} would be the same but the values for Newton and cosmological constants would be different for each possible $\mu$.}
\begin{eqnarray}
\theta(x)&\longmapsto&\mu\epsilon_{ab}e^a(X)e^b(X)\;,\nonumber\\
A(x)&\longmapsto&\frac{1}{2\mu}\epsilon_{ab}\omega^{ab}(X)\;,\label{map1}
\end{eqnarray}
provides the desired result for $[\mu]=1$. In \eqref{map1}, $x\in\mathbb{R}^2$ and $X\in\mathrm{M}$. Along with \eqref{map1}, the gauge transformation \eqref{gt1} can be mapped to \eqref{gt2} in two steps: First, $\alpha(x)\longmapsto\epsilon_{ab}\alpha^{ab}(X)$; Second, since $\theta$ is gauge invariant, so must be the \emph{rhs} of the first map in \eqref{map1}; Therefore, $e^a$ must be an $SO(2)$ covariant quantity.

At the level of the action, the map \eqref{map1} induces a Lovelock-Cartan gravity of the form
\begin{equation}
S_{sLC}=\int\frac{1}{8\pi G}\epsilon_{ab}\left(R^{ab}-\frac{\Lambda^2}{2}e^ae^b\right)\;,\label{grav2}
\end{equation}
with\footnote{It is worth mention that, for $\mu=a\kappa$ with $a\in\mathbb{R}^*_+$, Newton`s constant in \eqref{param0} is fixed to $\dfrac{1}{4\pi(2a+1)}$. This is consistent with the fact that $[G]=0$.}
\begin{equation}
 G=\frac{1}{4\pi}\frac{\mu}{\left(\kappa+2\mu\right)}\;,\;\;\Lambda^2=4\frac{\mu^3}{\left(\kappa+2\mu\right)}\;.\label{param0}
\end{equation}
Equations \eqref{param0} can be solved for $\mu$, namely
\begin{equation}
    \mu=\frac{1}{\sqrt{\pi G}}\frac{\Lambda}{2}\;.\label{param0a}
\end{equation}

Up to a minus sign in the cosmological constant, the action \eqref{grav2} is totally equivalent to the action \eqref{grav1}. The problem with the present proposal \eqref{map1} is that at one side we have only three degrees of freedom (from $A\equiv(A_0,A_1)$ and $\theta\equiv(\theta_{01}$)) and on the other side we have six degrees of freedom ($e^a\equiv(e^0_0,e^0_1,e^1_0,e^1_1)$ and $\omega^{ab}\equiv(\omega^{01}_0,\omega^{01}_1)$). Hence, to the map work in both ways, we need extra fields to account for the extra degrees of freedom.

\subsection{Improved map}

To have the right number of degrees of freedom we introduce the following matter action, to be added to electrodynamical action \eqref{ed2},
\begin{equation}
    S_m=\int\;d\left[\left(\bar{\varphi}\varphi-\bar{\zeta}\zeta\right)+\left(\bar{w}w-\bar{z}z\right)\right]\;,\label{sm}
\end{equation}
with $(\bar{\varphi},\varphi)$ and $(\bar{w},w)$ being real bosonic fields while $(\bar{\zeta},\zeta)$ and $(\bar{z},z)$ are hermitian fermionic fields. The fields $(\bar{\varphi},\bar{\zeta},\bar{w},\bar{z})$ are 0-forms while $(\varphi,\zeta,w,z)$ are 1-forms. The mass dimension of all extra fields are $[\bar{\varphi}]=[\bar{\zeta}]=[\bar{w}]=[\bar{z}]=0$ and $[\varphi]=[\zeta]=[w]=[z]=1$. The fact that we introduced integer spin fermions must be clarified. The reason is that we are aiming the quantization of the  model via BRST framework. Indeed, the auxiliary fields form a BRST octet system and do not affect the physical content of the model \cite{Piguet:1995er}. In fact, under BRST symmetry, the new auxiliary fields transform as
\begin{eqnarray}
 s\bar{\zeta}&=&\bar{\varphi}\;,\nonumber\\
s\bar{\varphi}&=&0\;,\nonumber\\
s\varphi&=&\zeta\;,\nonumber\\
s\zeta&=&0\;,\label{brs2a}
\end{eqnarray}
and
\begin{eqnarray}
 s\bar{z}&=&\bar{w}\;,\nonumber\\
s\bar{w}&=&0\;,\nonumber\\
sw&=&z\;,\nonumber\\
sz&=&0\;.\label{brs2b}
\end{eqnarray}
And the auxiliary action $S_m$ can be written as a BRST exact quantity,
\begin{equation}
    S_m=-s\int\;d\left(\bar{\zeta}\varphi+\bar{z}w\right)\;.\label{sm1}
\end{equation}
Thus, from the BRST point of view \cite{Piguet:1995er} we are not introducing any physical degree of freedom. Moreover, action \eqref{sm} is a surface term, only trivially affecting the boundary of $\mathbb{R}^2$. For completeness and further use, all quantum numbers of the fields on the electromagnetic side are displayed in Table \ref{table1}. 
\begin{table}[ht]
\centering
\begin{tabular}{|c|c|c|c|c|c|c|c|c|c|c|c|}
	\hline 
$\Phi$ & $A$ & $\theta$ & $c$ & $\bar{\varphi}$ & $\varphi$ & $\bar{\zeta}$ & $\zeta$ & $\bar{w}$ & $w$ & $\bar{z}$ & $z$ \\
	\hline 
D & $0$ & $1$ & $0$ & $0$ & $1$ & $0$ & $1$ & $0$ & $1$ & $0$ & $1$ \\ 
G & $0$ & $0$ & $1$ & $0$ & $0$ & $-1$ & $1$ & $0$ & $0$ & $-1$ & $1$\\
F & $1$ & $2$ & $0$ & $0$ & $1$ & $0$ & $1$ & $0$ & $1$ & $0$ & $1$\\
\hline 
\end{tabular}
\caption{Quantum numbers of the fields in the electromagnetic theory. We clarify that, $\Phi$ stands for the fields, D for the field dimension, G for the field ghost number, and F for the field form rank.}
\label{table1}
\end{table}

The full action to start with is then 
\begin{equation}
    S_{em}=S_e+S_m\;.\label{sem}
\end{equation}
It is easy to see that we have extra degrees of freedom at the side of electrodynamics, a total of fifteen against the six degrees of freedom at the gravity side. The consequence is that we will end up with extra degrees of freedom at the target gravity theory. In fact, a possible map accounting for this reads
\begin{eqnarray}
(\bar{\varphi}(x),\varepsilon\bar{\zeta}(x))\equiv\bar{\Phi}^a(x)&\longmapsto&\sigma^a(X)\;,\nonumber\\
(\varphi(x),-\varepsilon^{-1}\zeta(x))\equiv\Phi^a(x)&\longmapsto&\mu\phi^a_{\phantom{a}b}(X)e^b(X)\;,\nonumber\\
\label{map2}
\end{eqnarray}
and
\begin{eqnarray}
\bar{w}(x)&\longmapsto&\bar{\eta}(X)\;,\nonumber\\
\bar{z}(x)&\longmapsto&\bar{\chi}(X)\;,\nonumber\\
w(x)&\longmapsto&\mu\eta_a(X)e^a(X)\;,\nonumber\\
z(x)&\longmapsto&\mu\chi_a(X)e^a(X)\;,
\label{map3}
\end{eqnarray}
with $\sigma^a$, $\phi^a_{\phantom{a}b}$, $\bar{\eta}$, and $\eta^a$ are real bosonic fields while $\bar{\chi}$ and $\chi^a$ are hermitian ghost scalar fields. Moreover, the field $\phi^a_{\phantom{a}b}$ is assumed to be a Lorentz-algebra-valued field, therefore accounting for only one degree of freedom. Furthermore, all new fields are dimensionless. The maps \eqref{map2} maps 0-forms and 1-forms in the electromagnetic side into algebra-valued 0-forms in the gravity side. Such subtlety is required because the maps must not be $SO(2)$ invariant, such as \eqref{map1} and \eqref{map3}. Otherwise, we would still get redundancies and the full set of maps would be not invertible. In fact, the Jacobian of the maps \eqref{map1}, \eqref{map2} and \eqref{map3} can be calculated. The result, up to a normalization factor, is simply given by\footnote{See Appendix \ref{Ap1} for more details.}
\begin{equation}
    J=\mu^4\mathrm{det}^{3/2}(\phi^a_{\phantom{a}b})\mathrm{e}^3\;,\label{jac1}
\end{equation}
with $\mathrm{e}=\mathrm{det}(e^a_\mu)$.

Employing the maps \eqref{map1}, \eqref{map2} and \eqref{map3}, the action $S_m$ is mapped into
\begin{eqnarray}
S_{int}&=&\mu\int\;d\left(\sigma_a\phi^a_{\phantom{a}b}e^b+\bar{\eta}\eta_ae^a-\bar{\chi}\chi_ae^a\right)\nonumber\\
&=&\mu\int\left[D\sigma_a\phi^a_{\phantom{a}b}e^b+\sigma_a(D\phi^a_{\phantom{a}b}e^b+\phi^a_{\phantom{a}b}T^b)+d\bar{\eta}\eta_ae^a+\bar{\eta}\left(D\eta_ae^a+\eta_aT^a\right)+\right.\nonumber\\
&-&\left.d\bar{\chi}\chi_ae^a+\bar{\chi}\left(D\chi_ae^a-\chi_aT^a\right)\right]\;.\label{sm1a}
\end{eqnarray}
It is quite convenient to redefine the field $\phi^{ab}$ as 
\begin{equation}
    \phi_{ab}=\epsilon_{ab}\phi\;,\label{red1}
\end{equation}
so, action \eqref{sm1a} reads
\begin{equation}
    S_{int}=\mu\int\;d\left(\epsilon_{ab}\sigma^ae^b\phi+\bar{\eta}\eta_ae^a-\bar{\chi}\chi_ae^a\right)\;.\label{sm3}
\end{equation}
This action is, again, a BRST exact quantity,
\begin{equation}
    S_{int}=-\mu s\int\;d\left(\epsilon_{ab}\xi^ae^b\phi-\bar{\chi}\eta_ae^a\right)\;,\label{sm1b}
\end{equation}
if we define,
\begin{eqnarray}
 s\xi^a&=&\sigma^a+c^a_{\phantom{a}b}\xi^b\;,\nonumber\\
 s\sigma^a&=&c^a_{\phantom{a}b}\sigma^b\;,\nonumber\\
 s\bar{\chi}&=&\bar{\eta}\;,\nonumber\\
 s\bar{\eta}&=&0\;,\nonumber\\
 s\eta^a&=&\chi^a+c^a_{\phantom{a}b}\eta^b\;,\nonumber\\
 s\chi^a&=&c^a_{\phantom{a}b}\chi^b\;,\nonumber\\
 s\phi&=&0\;.\label{brs2c}
\end{eqnarray}
Therefore, the resulting gravity theory is finally given by
\begin{eqnarray}
    S_{grav}&=&S_{sLC}+S_{int}\nonumber\\
    &=&\int\frac{1}{8\pi G}\epsilon_{ab}\left(R^{ab}-\frac{\Lambda^2}{2}e^ae^b\right)+\mu\int\;d\left(\epsilon_{ab}\sigma^ae^b\phi+\bar{\eta}\eta_ae^a-\bar{\chi}\chi_ae^a\right)\;,
    \label{grav3}
\end{eqnarray}
which is equivalent to \eqref{grav2} up to a BRST trivial boundary term.

For completeness, the quantum numbers of the fields on the gravity theory are collected in Table \ref{table2}.
\begin{table}[ht]
\centering
\begin{tabular}{|c|c|c|c|c|c|c|c|c|c|c|}
	\hline 
$\Phi$ & $\omega^a_{\phantom{a}b}$ & $e^a$ & $c^a_{\phantom{a}b}$ & $\bar{\eta}$ & $\eta^a$ & $\phi^a_{\phantom{a}b}$ & $\bar{\chi}$ & $\chi^a$ & $\sigma$ \\
	\hline 
D & $1$ & $0$ & $0$ & $0$ & $0$ & $0$ & $0$ & $0$ & $0$ \\ 
G & $0$ & $0$ & $1$ & $0$ & $0$ & $0$ & $-1$ & $1$ & $0$ \\
F& $1$ & $1$ & $0$ & $0$ & $0$ & $0$ & $0$ & $0$ & $0$\\
\hline 
\end{tabular}
\caption{Quantum numbers of the fields in the gravity theory.}
\label{table2}
\end{table}

\section{Electrodynamics-gravity equivalence at quantum level}\label{QV}

The same map can be performed at quantum level. Thence, two-dimensional electrodynamics can be mapped into two-dimensional quantum gravity. Let us start with the generating functional of the electromagnetic theory
\begin{equation}
    Z_{qed}=\int[\mathcal{D}\Phi_{ed}]e^{-S_{ed}-S_m}\;,\label{zem1}
\end{equation}
with $[\mathcal{D}\Phi_{ed}]=\mathcal{D}A\mathcal{D}\theta\mathcal{D}\bar{\varphi}\mathcal{D}\phi\mathcal{D}\bar{\zeta}\mathcal{D}\zeta\mathcal{D}\bar{w}\mathcal{D}w\mathcal{D}\bar{z}\mathcal{D}z$ being the functional measure. Applying the maps \eqref{map1}, \eqref{map2} and \eqref{map3} and making use of the Jacobian \eqref{jac1}, one easily achieves
\begin{equation}
    Z_{grav}=\int[\mathcal{D}\Phi_{grav}]\mu^4\mathrm{det}^{3/2}(\phi^a_{\phantom{a}b})\mathrm{e}^3e^{-S_{grav}}\;,\label{zgrav1}
\end{equation}
with $[\mathcal{D}\Phi_{grav}]=\mathcal{D}\omega\mathcal{D}e\mathcal{D}\sigma\mathcal{D}\phi\mathcal{D}\bar{\eta}\mathcal{D}\eta\mathcal{D}\bar{\chi}\mathcal{D}\chi$. The determinants can be localized in the usual way. First, we consider
\begin{eqnarray}
\mu^2\mathrm{e}^3&=&\mu^2\mathrm{e}^4\mathrm{e}^{-1}\nonumber\\
&=&\int\mathcal{D}\bar{Y}\mathcal{D}Y\mathcal{D}Z\exp{\left[-\mu^2\int d^2X\left(\bar{Y}_ae^a_\mu Y^\mu+\frac{1}{2}Z_ae^a_\mu Z^\mu\right)\right]}\;,\nonumber\\
\label{det1}
\end{eqnarray}
with $\bar{Y}^a$ and $Y^\mu$ being ghost fields while $Z_a$ and $Z^\mu$ are bosonic fields. Moreover, the following relations are valid
\begin{eqnarray}
Z^a&=&e^a_\mu Z^\mu\;,\nonumber\\
Y^a&=&e^a_\mu Y^\mu\;,\label{yz1}
\end{eqnarray}
accounting for the correct number of degrees of freedom. Thus, in form notation, \eqref{det1} can be written as
\begin{equation}
    \mu^2\mathrm{det}^3(e^a_\mu)=\exp\left[-\mu^2\int\left(\bar{Y}_aY^a+\frac{1}{2}Z_aZ^a\right)\epsilon_{ab}e^ae^b\right]\;.\label{det1a}
\end{equation}

The same trick can be done for the other determinant in \eqref{zgrav1},
\begin{eqnarray}
\mu^2\mathrm{det}^{3/2}(\phi^a_{\phantom{a}b})&=&\mu^2\mathrm{det}^4(\phi^a_{\phantom{a}b})\mathrm{det}^{-1/2}(\phi^a_{\phantom{a}b})\nonumber\\
&=&\int\mathcal{D}\bar{W}\mathcal{D}W\mathcal{K}\exp\left[-\mu^2\int\left(\bar{W}_aW^b+\tilde{K}_aK^b\right).\phi^a_{\phantom{a}b}\epsilon_{cd}e^ce^d\right]\;,\label{det2}
\end{eqnarray}
with $\bar{W}^a$ and $W^a$ being 0-form ghost fields while $\tilde{K}_a$ and $K^a$ are 0-forms bosonic fields. Moreover, $\tilde{K}_a=\frac{1}{2}\epsilon_{ab}K^b$. We notice that $\tilde{K}_a\phi^a_{\phantom{a}b}K^b=2\phi K_0K^0$ with $\phi=\phi^0_{\phantom{0}1}$. Therefore, defining $K_0=\sigma/2$ and employing \eqref{red1} again, \eqref{det2} becomes
\begin{eqnarray}
\mu^2\mathrm{det}^{3/2}(\phi^a_{\phantom{a}b})&=&\int\mathcal{D}\bar{W}\mathcal{D}W\mathcal{D}\sigma\exp\left[-\mu^2\int\left(\epsilon_{ab}\bar{W}^aW^b+\frac{1}{2}\sigma^2\right)\phi\epsilon_{cd}e^ce^d\right]\;,\nonumber\\
&=&\int\mathcal{D}\bar{W}\mathcal{D}W\mathcal{D}\sigma\exp\left[-\mu^2\int\left(2\bar{W}_aW_be^ae^b+\frac{1}{2}\sigma^2\epsilon_{cd}e^ce^d\right)\phi\right]\;,\nonumber\\
\label{det2a}
\end{eqnarray}
Finally, the gravitational generating functional reads
\begin{equation}
    Z_{grav}=\int[\mathcal{D}\Phi_{grav}]e^{-S_{qg}}\;,\label{zgrav2}
\end{equation}
with
\begin{equation}
S_{qg}=S_{grav}+\mu^2\int\left(\bar{Y}_aY^a+\frac{1}{2}Z_aZ^a+\frac{1}{2}\sigma^2\phi\right)\epsilon_{ab}e^ae^b+2\mu^2\int\bar{W}_aW_be^ae^b\phi\;,\label{qg1}
\end{equation}
and $[\mathcal{D}\Phi_{grav}]$ incorporates all extra fields coming from the determinants.

The extra terms coming from the determinants in \eqref{qg1} can also be cast in a BRST exact form. Calling the action coming from the determinants $S_{aux}$, it reads
\begin{equation}
S_{aux}=\mu^2s\int\left[\left(\bar{Y}_aZ^a+\frac{1}{2}\bar{\sigma}\sigma\phi\right)\epsilon_{ab}e^ae^b+2\bar{W}_aM_be^ae^b\phi\right]\;,\label{aux1}
\end{equation}
with
\begin{eqnarray}
s\bar{\sigma}&=&\sigma\;,\nonumber\\
s\sigma&=&0\;,\nonumber\\
s\bar{Y}^a&=&Z^a\;,\nonumber\\
sZ^a&=&0\;,\nonumber\\
s\bar{W}^a&=&c^a_{\phantom{a}b}\bar{W}^b\;,\nonumber\\
sM^a&=&W^a+ c^a_{\phantom{a}b}M^b\;,\nonumber\\
sW^a&=&c^a_{\phantom{a}b}W^b\;.\label{brst4}
\end{eqnarray}
Thence, all physical content are contained in action \eqref{grav2}, as it should be since we have not introduced any extra physical degree of freedom in the electromagnetic theory.

To end this section, we display the quantum numbers of the new fields in Table \ref{table3}.
\begin{table}[ht]
\centering
\begin{tabular}{|c|c|c|c|c|c|c|c|c|}
	\hline 
$\Phi$ & $\bar{Y}^a$ & $Y^a$ & $Z^a$ & $\bar{W}^a$ & $W^a$ & $M^a$ & $\bar{\sigma}$ & $\sigma$ \\
	\hline 
D & $0$ & $0$ & $0$ & $0$ & $0$ & $0$ & 0 & 0\\ 
G & $-1$ & $1$ & $0$ & $-1$ & $1$ & $0$ & $-1$ & $0$  \\
F & $0$ & $0$ & $0$ & $0$ & $0$ & $0$ & $0$ & $0$ \\
\hline 
\end{tabular}
\caption{Quantum numbers of the localizing fields.}
\label{table3}
\end{table}

\section{Conclusions}\label{conc}

In this paper, we have formally developed a map from two-dimensional electrodynamics (action \eqref{ed1}) to two-dimensional gravity in the first order formalism (action \eqref{grav2}). To account for the different number of degrees of freedom in both theories, we introduced a set of auxiliary matter fields in such a way that the physical content in both theories are not affected - the auxiliary matter action is actually a BRST exact boundary term. Moreover, the respective mapped matter action is also a BRST exact boundary term. The map between theories is, in fact, not trivial. This non-triviality is evident in the computation of the Jacobian of the map, which, at quantum level, requires extra localizing fields at the gravity side of the mapping. However, the action containing such fields is also BRST exact. Thus, the Jacobian actually does not affect the physical content of the theory. 

We have established the basis to describe two-dimensional electrodynamics as a two-dimensional geometrodynamical theory. Nevertheless, much can be done in the future. For instance, the inclusion of non-trivial charged matter fields such as scalars and spinors would be the natural pathway to describe two-dimensional condensed matter systems\footnote{External matter was not considered here. However, its inclusion would be immediate since the coupling between the electromagnetic potential and an external current is given by $jA$ with $j$ being a 1-form describing the charged source. Clearly, it leads to a gravity coupling of the form $j_{ab}\omega^{ab}$.}. Another approach to attack two-dimensional systems would be considering four-dimensional electrodynamics followed by a suitable dimensional reduction to two dimensions and then, to perform the map. Such approach would be more realistic for actual physical systems. Moreover, The BRST approach for gauge fixing can be studied in the pure two-dimensional analysis performed in the present work or in the two-dimensional theory obtained from the dimensional reduction of four-dimensional theories. In fact, gauge fixing is a point that deserves extra discussion.

First of all, we must mention that gauge fixing is a necessary step in quantization of gauge theories if one needs to employ perturbative methods for actual computations. However, to compute observables (gauge invariants), no gauge fixing is needed \cite{Neuberger:1986vv,Neuberger:1986xz}. Thence, although beyond the scope of the present work, the qed-gravity correspondence here developed is ready to provide the expectation values of the observables of two-dimensional QED or 2-dimensional gravity described by $S_{qg}$. A direct possibility of gauge fixing is to consider the final action \eqref{qg1} and fix the gauge by considering the full BRST symmetry encompassing the BRST transformations associated to the $SO(2)$ gauge symmetry and diffeomorphism symmetries \cite{Baulieu:1984iw,Baulieu:1984pf,Moritsch:1993eg}. Another way is to fix the gauge in the QED side for the gauge field $A$. Due to the second relation in \eqref{map1}, a gauge fixing for $\omega$ is induced. Thus, fixing the $U(1)$ symmetry in the QED theory induces a gauge fixing for the $SO(2)$ symmetry of the gravity side. However, in the map \eqref{map1}, more degrees of freedom are introduced and the presence of the zweibein $e$, gluing the dynamics of the fields to spacetime, induces dynamics to the symmetries of spacetime, \emph{i.e.}, diffeomorphisms. Hence, I understand that extra gauge fixing must be imposed for the zweibein. To actually perform such gauge fixing, a possibility is to employ the BRST approach previously mentioned \cite{Baulieu:1984iw,Baulieu:1984pf,Moritsch:1993eg}. Another possibility is to consider the conformal gauge fixing developed by Polyakov in \cite{Polyakov:1987zb}, properly adapted to the first order formalism.

\section{Appendix: Computation of the Jacobian}\label{Ap1}

To compute the Jacobian of transformations \eqref{map1} and \eqref{map2} we need to construct the Jacobian matrix. For the map \eqref{map1}, the non-vanishing components are
\begin{eqnarray}
\frac{\delta\theta}{\delta e^a}&=&2\mu\epsilon_{ab}e^b\;,\nonumber\\
\frac{\delta A}{\delta\omega^{ab}}&=&\frac{1}{2\mu}\epsilon_{ab}\;.\label{jac2}
\end{eqnarray}
For the map \eqref{map2}, the only non-vanishing components are
\begin{eqnarray}
\frac{\delta\bar{\Phi}^a}{\delta\sigma^b}&=&\delta^a_b\;,\nonumber\\
\frac{\delta\Phi^a}{\delta e^a}&=&\mu\phi^a_{\phantom{a}b}\;,\nonumber\\
\frac{\delta\Phi^a}{\delta\phi^b_{\phantom{b}c}}&=&\delta^a_be^c\;.\label{jac3}
\end{eqnarray}
For the map \eqref{map3}, the non-vanishing components are
\begin{eqnarray}
\frac{\delta\bar{w}}{\delta\bar{\eta}}&=&1\;,\nonumber\\
\frac{\delta\bar{z}}{\delta\bar{\chi}}&=&1\;,\nonumber\\
\frac{\delta w}{\delta e^a}&=&\mu\eta_a\;,\nonumber\\
\frac{\delta w}{\delta\eta_a}&=&\mu e^a\;,\nonumber\\
\frac{\delta z}{\delta e^a}&=&-\mu\chi_a\;,\nonumber\\
\frac{\delta z}{\delta\chi_a}&=&\mu e^a\;.\label{jac4}
\end{eqnarray}
The components in \eqref{jac2}, \eqref{jac3} and \eqref{jac4}, together with all vanishing components, form a $15\times15$ matrix, namely $J_0$. This matrix can be immediately reduced to a $9\times9$ matrix, named $J_1$, due to the trivial relations in \eqref{jac2}, \eqref{jac3} and \eqref{jac4}. Thus, one can readly show that
\begin{equation}
    \mathrm{det}J_0=\frac{1}{\mu^2}\mathrm{det}J_1\;,\label{jac5}
\end{equation}
where $J_1$ can be arranged as
\begin{equation}
  \frac{J_1}{\mu}=
  \begin{bmatrix}
    0 & 0 & \phi & 0 & \eta_0 & 0 & \chi_0 & 0 & e^1_1 \\
    0 & 0 & 0 & \phi & 0 & \eta_0 & 0 & \chi_0 & -e^1_0 \\
    \phi & 0 & 0 & 0 & \eta_1 & 0 & \chi_1 & 0 & -e^0_1 \\
    0 & \phi & 0 & 0 & 0 & \eta_1 & 0 & \chi_1 & e^0_0 \\
    e^1_0 & e_1^1 & e^0_0 & e^0_1 & 0 & 0 & 0 & 0 & 0 \\
    0 & 0 & 0 & 0 & e^0_0 & e^0_1 & 0 & 0 & 0 \\
    0 & 0 & 0 & 0 & e^1_0 & e^1_1 & 0 & 0 & 0 \\
    0 & 0 & 0 & 0 & 0 & 0 & e^0_0 & e^0_1 & 0 \\
     0 & 0 & 0 & 0 & 0 & 0 & e^1_0 & e^1_1 & 0 \\
    \end{bmatrix}\label{jac6}
\end{equation}
where $\phi=\phi^0_{\phantom{0}1}$. The determinant of $J_1$ can be systematically computed within Laplace's method. After a long but straightforward computation, the result is given by
\begin{equation}
\mathrm{det}J_1=\mu^{9} \mathrm{det}(\phi^a_{\phantom{a}b})^{3/2} \mathrm{e}^3\;.\label{jac7}
\end{equation}
Thus,
\begin{equation}
    \mathrm{det}J_0=4\mu^7\mathrm{det}(\phi^a_{\phantom{a}b})^{3/2} \mathrm{e}^3\;,\label{jac8}
\end{equation}
which, up to a normalization factor, agrees with \eqref{jac1}.

\section*{acknowledgments}
This study was financed in part by The Coordena\c c\~ao de Aperfei\c coamento de Pessoal de N\'ivel Superior - Brasil (CAPES) - Finance Code 001.

\bibliography{BIB}

\providecommand{\href}[2]{#2}\begingroup\raggedright\begin{thebibliography}{10}

\bibitem{CastroNeto:2007fxn}
A.~H. Castro~Neto, F.~Guinea, N.~M.~R. Peres, K.~S. Novoselov, and A.~K. Geim,
  ``{The electronic properties of graphene}''.
  \href{http://dx.doi.org/10.1103/RevModPhys.81.109}{{\em Rev. Mod. Phys.}
  {\bfseries 81} (2009) 109--162}.

\bibitem{Farajollahpour:2019kwj}
T.~Farajollahpour, Z.~Faraei, and S.~A. Jafari, ``{Solid-state platform for
  space-time engineering: The 8Pmmn borophene sheet}''.
  \href{http://dx.doi.org/10.1103/PhysRevB.99.235150}{{\em Phys. Rev. B}
  {\bfseries 99} no.~23, (2019) 235150}.

\bibitem{Cahangirov:2009kwj}
S.~Cahangirov, M.~Topsakal, E.~Akturk, H.~Sahin, and C.~S., ``{Two- and
  One-Dimensional Honeycomb Structures of Silicon and Germanium}''.
  \href{http://dx.doi.org/10.1103/PhysRevLett.102.236804}{{\em Phys. Rev.
  Lett.} {\bfseries 102} (2009) 236804}.

\bibitem{Staruszkewicz:1967xxx}
A.~Staruszkewicz, ``{Fokker Action Principle and the Hyperbolic Motion}''.
  \href{http://dx.doi.org/10.1119/1.1974119}{{\em Am. J. Phys.} {\bfseries 35}
  (1967) 437}.

\bibitem{Bialynicki-Birula:1971akx}
I.~Bialynicki-Birula, ``{Classical electrodynamics in two dimensions - exact
  solution}''. \href{http://dx.doi.org/10.1103/PhysRevD.3.864}{{\em Phys. Rev.
  D} {\bfseries 3} (1971) 864--866}.

\bibitem{Kosyakov:1999np}
B.~P. Kosyakov, ``{Exact solutions of classical electrodynamics and the
  Yang-Mills-Wong theory in even-dimensional space-time}''.
  \href{http://dx.doi.org/10.1007/BF02557347}{{\em Theor. Math. Phys.}
  {\bfseries 119} (1999) 493--505}.

\bibitem{Kosyakov:2007qc}
B.~P. Kosyakov, \href{http://dx.doi.org/10.1007/978-3-540-40934-2}{{\em
  {Introduction to the classical theory of particles and fields}}}.
\newblock Springer, Berlin, Heidelberg, 2007.
\newblock
  \url{http://www.cambridge.org/uk/catalogue/catalogue.asp?isbn=0521545889}.

\bibitem{Kosyakov:2007np}
B.~P. Kosyakov, ``{Is classical reality completely deterministic?}''.
  \href{http://dx.doi.org/10.1007/s10701-007-9185-x}{{\em Found. Phys.}
  {\bfseries 38} (2008) 76--88}.

\bibitem{Polyakov:1987zb}
A.~M. Polyakov, ``{Quantum Gravity in Two-Dimensions}''.
  \href{http://dx.doi.org/10.1142/S0217732387001130}{{\em Mod. Phys. Lett. A}
  {\bfseries 2} (1987) 893}.

\bibitem{Gross:1989vs}
D.~J. Gross and A.~A. Migdal, ``{Nonperturbative Two-Dimensional Quantum
  Gravity}''. \href{http://dx.doi.org/10.1103/PhysRevLett.64.127}{{\em Phys.
  Rev. Lett.} {\bfseries 64} (1990) 127}.

\bibitem{Witten:1989ig}
E.~Witten, ``{On the Structure of the Topological Phase of Two-dimensional
  Gravity}''. \href{http://dx.doi.org/10.1016/0550-3213(90)90449-N}{{\em Nucl.
  Phys. B} {\bfseries 340} (1990) 281--332}.

\bibitem{Chamseddine:1989dm}
A.~H. Chamseddine, ``{Two-dimensional quantum gravity. A mini review}''.
  \href{http://dx.doi.org/10.1016/0920-5632(90)90601-P}{{\em Nucl. Phys. B
  Proc. Suppl.} {\bfseries 16} (1990) 579--580}.

\bibitem{Fukuma:1990jw}
M.~Fukuma, H.~Kawai, and R.~Nakayama, ``{Continuum Schwinger-dyson Equations
  and Universal Structures in Two-dimensional Quantum Gravity}''.
  \href{http://dx.doi.org/10.1142/S0217751X91000733}{{\em Int. J. Mod. Phys. A}
  {\bfseries 6} (1991) 1385--1406}.

\bibitem{Kazama:1992ex}
Y.~Kazama and H.~Nicolai, ``{On the exact operator formalism of two-dimensional
  Liouville quantum gravity in Minkowski space-time: A review}''. in {\em {4th
  Hellenic School on Elementary Particle Physics}}, pp.~753--789.
\newblock 9, 1992.

\bibitem{Achucarro:1986uwr}
A.~Achucarro and P.~K. Townsend, ``{A Chern-Simons Action for Three-Dimensional
  anti-De Sitter Supergravity Theories}''.
  \href{http://dx.doi.org/10.1016/0370-2693(86)90140-1}{{\em Phys. Lett. B}
  {\bfseries 180} (1986) 89}.

\bibitem{Witten:1988hc}
E.~Witten, ``{(2+1)-Dimensional Gravity as an Exactly Soluble System}''.
\href{http://dx.doi.org/10.1016/0550-3213(88)90143-5}{{\em Nucl. Phys.}
  {\bfseries B311} (1988) 46}.

\bibitem{Obukhov:1998gx}
{\relax Yu}.~N. Obukhov, ``{Gauge fields and space-time geometry}''.
  \href{http://dx.doi.org/10.1007/BF02557170}{{\em Theor. Math. Phys.}
  {\bfseries 117} (1998) 1308--1318}.
[Teor. Mat. Fiz.117,249(1998)].

\bibitem{Sobreiro:2007pn}
R.~F. Sobreiro and V.~J. Vasquez~Otoya, ``{Effective gravity from a quantum
  gauge theory in Euclidean space-time}''.
  \href{http://dx.doi.org/10.1088/0264-9381/24/20/003}{{\em Class. Quant.
  Grav.} {\bfseries 24} (2007) 4937--4953}.

\bibitem{Sobreiro:2011hb}
R.~F. Sobreiro, A.~A. Tomaz, and V.~J.~V. Otoya, ``{de Sitter gauge theories
  and induced gravities}''.
\href{http://dx.doi.org/10.1140/epjc/s10052-012-1991-4}{{\em Eur. Phys. J.}
  {\bfseries C72} (2012) 1991}.

\bibitem{Assimos:2013eua}
T.~S. Assimos, A.~D. Pereira, T.~R.~S. Santos, R.~F. Sobreiro, A.~A. Tomaz, and
  V.~J. Vasquez~Otoya, ``{From $SL(5,\mathbb{R})$ Yang-Mills theory to induced
  gravity}''.
\href{http://dx.doi.org/10.1142/S0218271817500870}{{\em Int. J. Mod. Phys.}
  {\bfseries D26} no.~08, (2017) 1750087}.

\bibitem{Assimos:2019yln}
T.~S. Assimos and R.~F. Sobreiro, ``{Constrained gauge-gravity duality in three
  and four dimensions}''.
  \href{http://dx.doi.org/10.1140/epjc/s10052-019-7552-3}{{\em Eur. Phys. J. C}
  {\bfseries 80} no.~1, (2020) 20}.

\bibitem{Assimos:2021eok}
T.~S. Assimos and R.~F. Sobreiro, ``Topological gauge-gravity equivalence:
  fiber bundle and homology aspects''. (2021).
\newblock \url{https://arxiv.org/pdf/2111.14675.pdf}. e-print: 2111.01672
  [hep-th].

\bibitem{Becchi:1975nq}
C.~Becchi, A.~Rouet, and R.~Stora, ``{Renormalization of Gauge Theories}''.
  \href{http://dx.doi.org/10.1016/0003-4916(76)90156-1}{{\em Annals Phys.}
  {\bfseries 98} (1976) 287--321}.

\bibitem{Tyutin:1975qk}
I.~V. Tyutin, ``{Gauge Invariance in Field Theory and Statistical Physics in
  Operator Formalism}''. (1975).
\newblock \url{https://arxiv.org/pdf/0812.0580.pdf}. LEBEDEV-75-39, e-print:
  0812.0580 [hep-th].

\bibitem{Piguet:1995er}
O.~Piguet and S.~P. Sorella, ``{Algebraic renormalization: Perturbative
  renormalization, symmetries and anomalies}''.
\href{http://dx.doi.org/10.1007/978-3-540-49192-7}{{\em Lect. Notes Phys.
  Monogr.} {\bfseries 28} (1995) 1--134}.

\bibitem{Utiyama:1956sy}
R.~Utiyama, ``{Invariant theoretical interpretation of interaction}''.
  \href{http://dx.doi.org/10.1103/PhysRev.101.1597}{{\em Phys. Rev.} {\bfseries
  101} (1956) 1597--1607}.

\bibitem{Kibble:1961ba}
T.~W.~B. Kibble, ``{Lorentz invariance and the gravitational field}''.
  \href{http://dx.doi.org/10.1063/1.1703702}{{\em J. Math. Phys.} {\bfseries 2}
  (1961) 212--221}.

\bibitem{Sciama:1964wt}
D.~W. Sciama, ``{The Physical structure of general relativity}''.
  \href{http://dx.doi.org/10.1103/RevModPhys.36.1103}{{\em Rev. Mod. Phys.}
  {\bfseries 36} (1964) 463--469}. [Erratum: Rev.Mod.Phys. 36, 1103--1103
  (1964)].

\bibitem{Lovelock:1971yv}
D.~Lovelock, ``{The Einstein tensor and its generalizations}''.
  \href{http://dx.doi.org/10.1063/1.1665613}{{\em J. Math. Phys.} {\bfseries
  12} (1971) 498--501}.

\bibitem{Mardones:1990qc}
A.~Mardones and J.~Zanelli, ``{Lovelock-Cartan theory of gravity}''.
  \href{http://dx.doi.org/10.1088/0264-9381/8/8/018}{{\em Class. Quant. Grav.}
  {\bfseries 8} (1991) 1545--1558}.

\bibitem{Zanelli:2005sa}
J.~Zanelli, ``{Lecture notes on Chern-Simons (super-)gravities. Second edition
  (February 2008)}''. in {\em {Proceedings, 7th Mexican Workshop on Particles
  and Fields (MWPF 1999): Merida, Mexico, November 10-17, 1999}}.
\newblock
2005.
\newblock

\bibitem{Zanelli:2012zz}
J.~Zanelli, ``{Introductory lectures on Chern-Simons theories}''.
  \href{http://dx.doi.org/10.1063/1.3678608}{{\em AIP Conf. Proc.} {\bfseries
  1420} no.~1, (2012) 11--23}.

\bibitem{Nakahara:2003nw}
M.~Nakahara, {\em {Geometry, topology and physics}}.
\newblock Boca Raton, USA: Taylor \& Francis (2003) 573 p., 2003.

\bibitem{Wald:1984rg}
R.~M. Wald,
  \href{http://dx.doi.org/10.7208/chicago/9780226870373.001.0001}{{\em {General
  Relativity}}}.
\newblock Chicago Univ. Pr., Chicago, USA,
1984.
\newblock

\bibitem{DeSabbata:1986sv}
V.~De~Sabbata and M.~Gasperini, {\em {INTRODUCTION TO GRAVITY}}.
\newblock Singapore, Singapore: World Scientific 346p, 1985.

\bibitem{Misner:1973prb}
C.~W. Misner, K.~S. Thorne, and J.~A. Wheeler, {\em {Gravitation}}.
\newblock W. H. Freeman, San Francisco, 1973.

\bibitem{Moritsch:1993eg}
O.~Moritsch, M.~Schweda, and S.~P. Sorella, ``{Algebraic structure of gravity
  with torsion}''. \href{http://dx.doi.org/10.1088/0264-9381/11/5/010}{{\em
  Class. Quant. Grav.} {\bfseries 11} (1994) 1225--1242}.

\bibitem{Baulieu:1984iw}
L.~Baulieu, ``{Anomalies and Gauge Symmetry}''.
  \href{http://dx.doi.org/10.1016/0550-3213(84)90060-9}{{\em Nucl. Phys. B}
  {\bfseries 241} (1984) 557--588}.

\bibitem{Baulieu:1984pf}
L.~Baulieu and J.~Thierry-Mieg, ``{Algebraic Structure of Quantum Gravity and
  the Classification of the Gravitational Anomalies}''.
  \href{http://dx.doi.org/10.1016/0370-2693(84)90946-8}{{\em Phys. Lett. B}
  {\bfseries 145} (1984) 53--60}.

\bibitem{Neuberger:1986vv}
H.~Neuberger, ``{NONPERTURBATIVE BRS INVARIANCE}''.
  \href{http://dx.doi.org/10.1016/0370-2693(86)90333-3}{{\em Phys. Lett. B}
  {\bfseries 175} (1986) 69--72}.

\bibitem{Neuberger:1986xz}
H.~Neuberger, ``{Nonperturbative {BRS} Invariance and the Gribov Problem}''.
  \href{http://dx.doi.org/10.1016/0370-2693(87)90974-9}{{\em Phys. Lett. B}
  {\bfseries 183} (1987) 337--340}.

\end{thebibliography}\endgroup
\bibliographystyle{utphys2}

\end{document}